\documentstyle[12pt,a4wide]{article}
\input epsf

\newcommand{\ssf}{\sans}
\newcommand{\news}{\setcounter{equation}{0}}
\newcommand{\be}{\begin{equation}}
\newcommand{\ee}{\end{equation}}
\newcommand{\R}{\hbox{\upright\rlap{I}\kern 1.7pt R}}
\newcommand{\Z}{\hbox{\upright\rlap{\ssf Z}\kern 2.7pt {\ssf Z}}}
\newcommand{\bea}{\begin{eqnarray}}
\newcommand{\eea}{\end{eqnarray}}
\newcommand{\alphabf}{\mbox{\boldmath$\alpha$}}
\newcommand{\betabf}{\mbox{\boldmath$\beta$}}
\newcommand{\gammabf}{\mbox{\boldmath$\gamma$}}

\font\upright=cmu10 scaled\magstep1
\font\sans=cmss12
\input mssymb

\begin{document}
\pagestyle{plain}
\title{\vskip -70pt
  \begin{flushright}
    {\normalsize DAMTP 97-33. hep-th/9704153.} 
  \end{flushright}
  \vskip 15pt
  {\bf \Large \bf SU(3) monopoles and their fields.} 
  \vskip 10pt}
 
\author{Patrick Irwin\thanks{Electronic address :
    P.W.Irwin@damtp.cam.ac.uk}\\
  {\normalsize {\sl Department of Applied Mathematics
      and Theoretical Physics,}}\\
  {\normalsize {\sl University of Cambridge, Silver St.,
      Cambridge CB3 9EW, U.K.}}}

\maketitle
\begin{abstract}
Some aspects of the fields of charge two SU(3) monopoles with
minimal symmetry breaking are discussed. A certain class of solutions look like
SU(2) monopoles embedded in SU(3) with a transition region or
``cloud'' surrounding the monopoles. For large cloud size the relative
moduli space metric splits as a direct product ${\cal AH}\times \R^4$
where ${\cal AH}$ is the Atiyah-Hitchin metric for SU(2) monopoles
and $\R^4$ has the flat metric. Thus the cloud is parametrised by
$\R^4$ which corresponds to its radius and SO(3) orientation.
We solve for the long-range fields in this
region, and examine the energy density and rotational moments of
inertia. The moduli space metric for these monopoles, given by Dancer, is
also expressed in a more explicit form.
\end{abstract}

\section{Introduction}
\news
\ \indent
In this paper we examine some aspects of charge two BPS monopoles in a 
gauge theory SU(3)
spontaneously broken by an adjoint Higgs to U(2). Their moduli space
was found by Dancer in \cite{3}, and subsequently studied in 
\cite{4,5,13}.
Magnetic monopoles in SU(2) 
gauge theory have been the focus of
considerable interest. Recently, there has been renewed
interest in monopoles of higher rank gauge groups mainly in relation
to electric-magnetic duality \cite{GL, LWY1, Connell, LWY3, GW}.
Substantial progress has been made in theories where the gauge symmetry
is broken to the maximal torus. However if the unbroken symmetry group
contains a non-Abelian component various complications arise and much
less is known.
For SU(3) monopoles with minimal symmetry breaking the
topological charge is specified by a single integer $k$.
But as pointed out in
\cite{GNO} there is a further gauge invariant classification of
monopoles, with the result that the charge $k$ moduli space is
stratified into different connected components. Charge one monopoles 
are given by embeddings of
the charge one SU(2) monopole. Their moduli space is given by
$\R^3\times S^3$, which is fibred over $S^2$ with fibre 
$\R^3 \times U(1)$. 
Points in $S^2$ label different possible embeddings of the SU(2) solution,
however it is well known that it is not possible to move along the
$S^2$ factor \cite{19, 20, 21} due to the non-Abelian nature of the long-range
fields (the long-range magnetic field does not commute with all the 
generators of the unbroken gauge group).

Charge two monopoles appear in two
distinct categories depending on whether or not the long-range fields
are Abelian.
Monopoles with non-Abelian long-range magnetic fields  are given by
embeddings of charge two SU(2) solutions, the moduli space is
of dimension ten (which is fibred over $S^2$ with fibre the charge 
two SU(2) moduli space). However if the long-range magnetic fields
are Abelian the
moduli space is more
complicated and has dimension twelve. This phenomenon where the charge
$k$ moduli space is stratified, with different components  having different
dimensions occurs for any gauge theory in which the unbroken subgroup
is non-Abelian \cite {Bow, Murray}. 

In the presence of a
monopole whose long-range fields are non-Abelian
it is not
possible in general to define a global color gauge transform which is 
orthogonal to the orbits of gauge transforms which are 1 at
spatial infinity (i.e. little gauge
transforms) \cite{19}. As Abouelsaood showed in \cite{20}, this is a
consequence of the result \cite{21}, which showed the impossibility of globally
defining (in any regular gauge) a full set of generators that commute 
with the Higgs field on a
two sphere at infinity when the monopole has long-range non-Abelian fields.
In the SU(3) 
theory this is so for embedded SU(2) monopoles. However, when
the long-range fields are Abelian such
transformations are possible and the moduli space gains a
normalisable action of SU(2). This
suggests that the moduli space for the larger strata would be of
dimension eleven rather than eight as for embedded SU(2) monopoles (ignoring
the $S^2$ fibre). However,
it is known that the moduli space is of dimension twelve \cite{Bow, Wein1}, 
and so an extra parameter appears  whose physical interpretation is
somewhat unclear. We shall see that in certain regions the monopoles
look like embedded SU(2) monopoles surrounded by a non-Abelian cloud
whose approximate radius can be attributed to this extra parameter.

It has been proved that the moduli space of monopoles for many gauge 
groups is in
1-1 correspondence with the moduli space of solutions to Nahm's
equations on a given interval with certain boundary
conditions \cite{Nahm}. This correspondence is known to be an isometry
for SU(2)
monopoles and SU(n+1)  monopoles broken to U(n), which includes the
present case \cite{NAK1, NAK2}. In \cite{3}, Dancer has constructed
the metric on the moduli space of solutions to Nahm's equations which 
is a twelve dimensional
hyperk\"{a}hler manifold with free, isometric actions of $R^3$, U(2),
and Spin(3). By \cite{NAK2} this gives the metric on the two
monopole moduli space. 

The Dancer manifold can be thought of as having a boundary which corresponds
to the space of embedded SU(2) monopoles (however the manifold with
the induced $L^2$ metric is
complete and the boundary is infinitely far away in metric distance). 
Like in the SU(2) case, the monopoles have a
well-defined center of mass and total U(1) phase. In \cite{3}
an implicit expression was given for the relative moduli space metric
which has isometry group SO(3)$\times$SO(3). Below, this
metric is expressed in terms of invariant one-forms
corresponding to the actions of each of the SO(3)'s and two other
parameters which roughly measure the monopoles' separation and how
``close'' the monopole configuration is to an embedded SU(2) monopole 
configuration. The expressions for the metric are very complicated and
are relegated to the Appendix. However for regions of the moduli space
which approach the boundary the metric simplifies into a direct
product of ${\cal AH}\times \R^4$. An interpretation
is that the configuration looks like an embedded
charge two SU(2) monopole surrounded by a ``cloud'' \cite{LWY1},
which is parametrised by its physical radius (which is related to the
inverse of the coordinate distance from
the boundary of the Dancer manifold) and its SO(3) orientation 
(residual gauge group
action). Evidence for this is given by solving for the long-range
Higgs and gauge fields in this region of the moduli space. 
Here, the long-range fields do indeed have this cloud where
the fields change from those of a non-Abelian 
embedded SU(2) monopole to the Abelian fall-off of these
monopoles. 
The long-range fields are obtained from a spherical
symmetry ansatz and so our results are only valid at a distance much larger
than the separation of the two monopoles. The kinetic energy obtained by 
varying the size of the cloud may be calculated and is shown to agree with the
kinetic energy deduced from the metric. The moments of inertia,
corresponding to the SO(3) gauge action, can be read off 
from the metric and they diverge as the cloud
spreads out to infinity.

The presence of the cloud can alter the possible types of 
monopole scattering from the SU(2) case.
For SU(2) monopoles if two monopoles collide head on, they
form a torus and then scatter at right angles. Here a
different outcome is possible \cite {5}. Incoming 
monopoles can collide, forming instantaneously a spherically 
symmetric monopole, but now
instead of scattering outwards they continue
to approach the embedded SU(2) torus. Due to the
conservation of kinetic energy of the incoming monopoles (and angular
momentum conservation) the SU(2) monopoles must scatter but in the
SU(3) case the kinetic energy is carried off by the cloud while the
monopoles' core is static in the limit of large time. 

An alternative description of this cloud was given by Lee, Weinberg
and Yi \cite{LWY1}. They
proposed that the moduli space of monopoles whose magnetic charge is Abelian
can be obtained as a limit of a monopole moduli space in a theory where the
gauge group is broken to its maximal torus. Our case, where SU(3) is
broken to U(2),
can be viewed as a limit of SU(3) broken to
U(1)$\times$U(1) as the Higgs expectation value is varied.
Then the Dancer moduli space would arise from a space of
three monopoles : two of the same type, and the third distinct and
becoming massless in the limit. As this monopole becomes massless its
core radius expands and eventually loses its identity as a monopole
and is manifested as a cloud surrounding the remaining two monopoles.

Section 2 is a review of SU(3) monopoles. For completeness we
discuss both cases, SU(3) broken to U(1)$\times$U(1) and
SU(3) broken to U(2). In section 3 we consider the charge two
monopoles where SU(3) is broken to U(2). In particular we 
solve for the long-range gauge
and Higgs fields in regions of the moduli space which are close to
the boundary of embedded SU(2) monopoles. In the Appendix the metric
given in \cite{3} is reexpressed in an explicit form.
\section{Review of SU(3) Monopoles}
\news
\ \indent
  We assume throughout that the Higgs
field is in the adjoint representation and we are in the BPS limit in
which the scalar potential is zero but a nonzero Higgs expectation
value is imposed
at spatial infinity as a boundary condition.
An SU(3) gauge theory can be broken by an adjoint Higgs mechanism
to either U(1)$\times$U(1) or U(2). Monopole solutions will occur
in either theory. The generators of SU(3)  may be chosen to be
two commuting operators $H_1$, $H_2$, together with ladder operators,
associated with the roots $\pm\alphabf$, $\pm\betabf$, $\pm(\alphabf+\betabf)$
(see figure 1), that obey 
\be
[H_i,E_{\gammabf}]=\gamma^iE_{\gammabf},\;\;\;\;\; [E_{\gammabf},
E_{-\gammabf}]=\gamma^iH_i\;,
\ee
for $\gammabf$ any root.
Following \cite{Wein1} we let $\Phi_{\infty}$ be the asymptotic
value of the Higgs field along the positive $x^3$-axis. Choose this to lie
in the Cartan subalgebra and this defines a vector $\bf{h}$ by   
$\Phi_{\infty}=\bf{h}.\bf{H}$.
If SU(3) is broken to U(1)$\times$U(1), all roots have nonzero inner
product with $\bf{h}$ and there is a unique set of simple roots with
positive inner product with $\bf{h}$. If SU(3) is broken to U(2)
then one of the roots, say $\betabf$, is perpendicular to $\bf{h}$. Now
there are two choices of simple roots with non-negative inner product
with $\bf{h}$; namely $(\alphabf,\betabf)$ and
$(\alphabf+\betabf,-\betabf)$. Figure 1 illustrates the two different
types of symmetry breaking.

\begin{figure}[ht]
\begin{center}
\leavevmode
\epsfxsize=13cm\epsffile{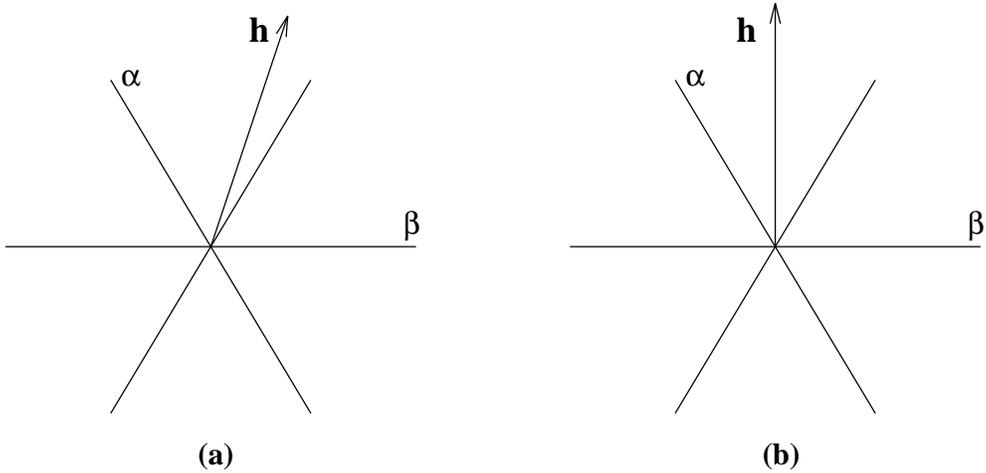}
\caption{(a) SU(3)$\rightarrow U(1)\times U(1)\;\;\;\;$(b) SU(3)$\rightarrow$ U(2)}
\end{center}
\end{figure}

For any finite energy solution, asymptotically 
\be
B_i=\frac{r_i}{4\pi r^3}G(\Omega)
\ee
$G(\Omega)$ is a covariantly constant element of the Lie algebra of
SU(3), and takes the value $G_0$ along the
positive $x^3$-axis. The Cartan subalgebra may be chosen so that 
$G_0=\bf{g}.\bf{H}$.
The quantisation of magnetic charge \cite{GNO,EW}, ensures that
$\bf{g}$ is of the form 
\be
{\bf{g}}=\frac{4\pi}{e}\left\{n\alphabf^*+m\betabf^*\right\}\;,\;\;\;\;\;\;
\alphabf^*=\frac{\alphabf}{\alphabf\cdot\alphabf}
\ee
where $e$ is the gauge coupling, and $n$, $m$ are non-negative integers.
We denote such a charge by $(n,m)$.

When SU(3) is broken to U(1)$\times$U(1) the topological
charges of the monopoles are determined by a pair of integers,
ie. the monopoles
can be charged with respect to either of the unbroken U(1)'s.
All BPS monopoles may
be thought of as superpositions of two fundamental monopoles given by
embeddings of the charge one SU(2) monopole \cite{Wein1}. 
To embed an SU(2) solution note that any root $\gammabf$ defines an
SU(2) subalgebra by
\bea
t^1(\gammabf)&=&(2{\gammabf\cdot\gammabf})^{-1/2}(E_{\gammabf}+E_{-\gammabf})\\
t^2(\gammabf)&=&-i(2{\gammabf\cdot\gammabf})^{-1/2}(E_{\gammabf}-E_{-\gammabf})\nonumber\\
t^3(\gammabf)&=&\gammabf^*\cdot{\bf{H}}\;.\nonumber
\eea

If $A_i^a$ and ${\Phi}^a$ is an SU(2) charge $n$ BPS solution with
Higgs expectation value $v$, then a monopole with magnetic charge  
\be
{\bf{g}}=\frac{4\pi n}{e}\gammabf^*
\ee
is given by \cite{Bais}
\bea
\Phi &=& \Phi^at^a +({\bf h}-\frac{{\bf h}\cdot
\gammabf}{\gammabf\cdot\gammabf}\gammabf)\cdot{\bf H}\\
{\bf A}_i &=& A_i^a t^a \nonumber\\
v &=& {\bf h}\cdot\gammabf\;\;.\nonumber
\eea

The two fundamental monopoles are obtained by embedding charge one
solutions along the simple
roots, $\alphabf$ and $\betabf$. Each fundamental monopole has four
zero modes, corresponding to its position and a U(1) phase.
Embedding along the root $\alphabf$ gives the (1,0) monopole
charged with respect to one of the U(1)'s. Similarly, one can embed 
along the root $\betabf$ to give the monopole (0,1)
charged with respect to the other U(1). Any BPS solution of
topological charge ($n$,$\,m$) can be viewed as a collection 
of $n$ $\,\alphabf$-monopoles and $m$  
$\,\betabf$-monopoles. The dimension of the $(n,\,m)$ moduli space is
$4(n+m)$. The ($n$,$\,0$) and ($0$,$\,n$)
moduli spaces will be copies of the charge $n$ SU(2) moduli space. 
The ($1$,$\,1$) moduli space was studied in \cite{Connell}, see also \cite{GL}.
It contains the spherically symmetric
solution obtained by embedding a single SU(2) monopole along the 
root $\alphabf +\betabf$. The relative moduli space is 
Taub-NUT (Newman-Unti-Tamburino) with positive mass
parameter. Because the monopoles are charged with respect to different
U(1)'s their interaction is simpler than for two SU(2) monopoles.
The other moduli spaces are known as spaces of holomorphic rational
maps from the two sphere into a flag manifold \cite{Hurt}. But their 
moduli space metrics are as yet unknown, although a conjectured
form for the (2,1) moduli space metric was given in \cite{Chalmers}.

For SU(3) broken to  U(2) the situation is quite different. Now the
vacuum expectation value of the Higgs field along the $x^3$-axis, ${\bf h}$, 
is perpendicular to one
of the roots, $\betabf$. Embedding an SU(2) solution using the SU(2)
subalgebra defined by the root $\betabf$, (2.4), now gives the
trivial zero solution so (0,1) solutions do not exist here. In
general, 
the quantization of magnetic charge is again
given by (2.3) but now $n$ is the only topological charge. Solutions
for a given value of $n$ exist only if $m\leq n$, and the $(n,m)$
moduli space is 
identical to the $(n,n-m)$ moduli space.

Two distinct charge one solutions are
given by embeddings along the roots $\alphabf$ and
$\alphabf+\betabf$, they are (1,0) and (1,1) respectively. The 
long-range magnetic field has a non-Abelian component ie., $G_0$ does not
commute with  the generator of the unbroken SU(2)
(${\bf{g}}\cdot\betabf\neq 0$) so it is not possible in general to
perform a  global gauge transform $g=e^{-\Gamma}$, taking
values in the unbroken U(2) at infinity which 
is orthogonal to little gauge
transforms \cite{19}, i.e. $D_iD_i \Gamma+[\Phi,[\Phi,\Gamma]]=0$. 
This is so only for gauge transforms generated by
$E_{\betabf},\,E_{-\betabf}$. Gauge transforms in the Cartan subalgbra
pose no such difficultly. However a linear combination of the Cartan
generators leaves the monopole invariant so 
only a global U(1) charge rotation remains 
($g=e^{-\chi\Phi}$) exactly as for SU(2) monopoles.
However there are more charge one solutions than this, because one may
act with the global SU(2) in the singular gauge,  
but one cannot move between these solutions dynamically
(the embedded
solutions based on the roots $\alphabf$ and $\alphabf+\betabf$ are
related in this fashion). This gives a
three dimensional family of solutions parametrised by $S^3$, and 
with translational 
invariance the space of solutions is of the form $\R^3\times S^3$
which may be viewed as a fibre bundle over $S^2$ with fibre
$\R^3\times$U(1).
However, the zero modes corresponding to the action of the global
unbroken SU(2) gauge group on the monopole (corresponding to the
$S^2$ factor in the moduli space) cannot satisfy $both$
the linearized BPS equation and orthogonality to little gauge
orbits. Only the U(1) factor corresponding to electric charge
rotations can do this. So, although the space of solutions is
$\R^3\times S^3$, it is clear that the usual procedure of 
finding the metric by calculating the $L^2$
norms of the zero modes satisfying $D_i\delta {\bf A}_i+[\Phi,\delta
\Phi]=0$ is not possible here.

For topological charge two solutions, either one can embed charge two SU(2)
solutions along the root $\alphabf$ or  $\alphabf+\betabf$, giving
(2,0), (2,2) respectively, or,
alternatively one may combine the two charge one solutions based on the
roots $\alphabf$ and $\alphabf+\betabf$ giving (2,1). In
the former case, again there is a long-range non-Abelian magnetic
field and the same problems as for charge one solutions are
present. The moduli space
will be the charge two SU(2) monopole space fibred over $S^2$. In the
latter case, the  magnetic charge is Abelian, (${\bf{g}}\cdot\betabf=0$),
so global color (SU(2)) gauge transforms are possible with finite
$L^2$ norm for the zero modes \cite{19}.
The moduli space acquires an action of SU(2), but as stated in the
introduction another parameter is present on the moduli space whose
appearance is somewhat surprising. 
As noticed in \cite{Bow} the parameter counting means these monopoles
cannot be interpreted as a superposition of any fundamental monopoles,
unlike the SU(3)$\rightarrow$U(1)$\times$U(1) case.
In the following we shall be studying SU(3) broken to U(2)
charge (2,1) monopoles which have an Abelian magnetic charge.
\section{The Asymptotic Fields}
\news
\ \indent
\begin{figure}
\begin{center}
\leavevmode
\epsfxsize=3in\epsffile{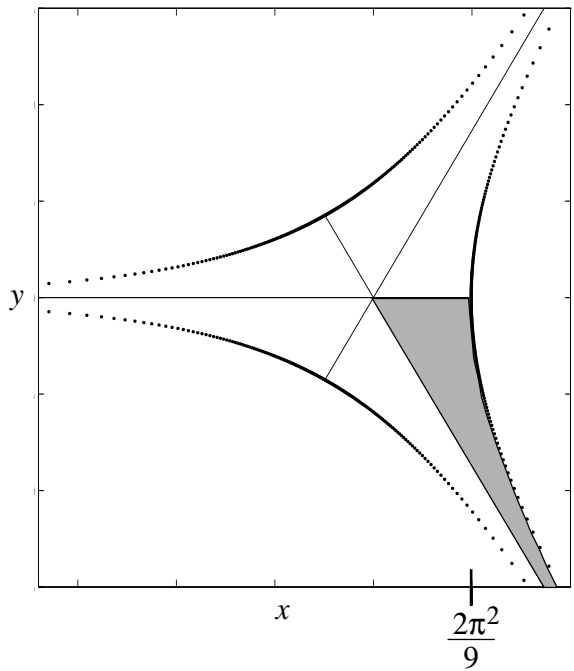}
\caption{The geodesic submanifold $Y$.}
\end{center}
\end{figure}
Nakajima and Takahasi have proved that the metric on the moduli space
of SU(2) monopoles and SU(n+1) monopoles broken to U(n)
is equivalent to the metric on a corresponding moduli space of
solutions of Nahm's equations \cite{NAK1,NAK2}.
This equivalence provides a direct method of finding metrics on
monopole moduli spaces, at least for low charges where Nahm's
equations can be solved. Using the
Hyperk\"{a}hler quotient construction, Dancer \cite{3}, has
constructed the metric on the 
moduli space of Nahm data corresponding to SU(3) broken to U(2)
charge (2,1) monopoles denoted $M^{12}$
and by Takahasi's proof on the equivalence of the the two metrics this
gives the monopole metric.
$M^{12}$ is a twelve dimensional manifold with commuting actions of
Spin(3) (rotations), U(2) (unbroken gauge group), and $\R^3$ 
(translations). Let $M^8$ be the quotient of $M^{12}$ by
$\R^3\times U(1)$ where U(1) is the center of U(2). $M^8$ has free
commuting actions of SO(3) and SU(2)$/\Z_2$. The
metric on $M^{12}$ is just the Riemannian product of the metric on
$M^8$ with the flat metric on $\R^3\times U(1)$ so the manifold $M^8$ 
describes the relative motion of the monopoles.  
$M^8$ may be
quotiented by the SU(2)$/\Z_2$ action to give a manifold denoted by
$N^5$ which has a non-free action of SO(3), corresponding to rotations.
In \cite{3}, an explicit expression was given for the metric on $N^5$
and an implicit expression for the metric on $M^8$. In the Appendix
the metrics on $N^5$ and $M^8$ are reexpressed in terms of coordinates
$D$, $k$, on the quotient space $N^5/$SO(3) and left-invariant 1-forms on
SO(3) corresponding to the action of SO(3) on $N^5$ and
SO(3)$\times$SU(2)$/\Z_2$ on $M^8$.

Here we are interested in the interpretation of the moduli space
coordinates in terms of the monopole fields. After the removal
of gauge freedom, spatial translations and rotations, we are left
with the two dimensional space $N^5/$SO(3). $N^5/$SO(3) may be
parametrised by $D$, $k$ with $0\leq k\leq 1$ and $0\leq D
<\frac{2}{3}K(k)$, where $K(k)$ denotes the first complete 
elliptic integral $K(k)=
\int^{\frac{\pi}{2}}_0(1-k^2\sin^2\theta\,)^{-1/2}d\,\theta$ \cite{3}.

Figure 2 depicts a totally geodesic submanifold of $M^8$, $\,Y$,
corresponding to monopoles which are
reflection symmetric (up to a gauge transform) in each of the three
Cartesian axes. $Y$ is composed of six regions each isomorphic 
to $N^5/$SO(3). The
metric and geodesic flow on $Y$ was studied in \cite{5}. In the shaded
region $x,\,y$ are given in terms of $D,\,k$ by
\be
x=(2-k^2)D^2\,,\qquad
y=-\sqrt{3}k^2D^2\;.
\ee
Similar formulae for $x,\,y$ in terms of $D,\,k$ exist in the other
regions of $Y$.
The origin, $x=y=0$ corresponds to the spherically symmetric monopole
for which the fields are known, \cite{4, BW}. The line segments
 $-\infty<x\leq 0$,$\,\,y=0$ and
$0\leq x<\frac{2}{9}\pi^2$,$\,\,y=0$
correspond to axially symmetric monopoles, denoted in \cite{4} as
hyperbolic and trigonometric  monopoles respectively, because the Nahm
data involves hyperbolic and trigonometric functions. In terms of $D$
and $k$ these are lines $k=1$ and $k=0$. These line segments together
correspond to a geodesic on $Y$. As $x$ varies from $-\infty$ to 
$\frac{2}{9}\pi^2$, two infinitely separated hyperbolic monopoles
approach each other, collide to form the spherically symmetric monopole, 
which then deforms to trigonometric monopoles and approaches the 
axially symmetric  
embedded  SU(2) monopole. In \cite{4}, for these cases,
the Higgs field was determined along the axis of axial symmetry.
The dotted boundaries of $Y$ [which is $D=\frac{2}{3}K(k)$ in
$N^5/$SO(3)] corresponds to embedded SU(2) monopoles. $Y$ is geodesically
complete and the boundary is infinitely far away in metric distance.

We want to understand the nature of the metric and the fields in
different regions of $Y$. From numerical evidence in \cite{4,5}, for
large $D$, $D$ represents the separation of the monopoles. The
asymptotic regions of Figure 2 (the legs of $Y$) correspond to 
well separated monopoles
along each of the three axes. The central region corresponds to 
configurations of coincident or closely separated monopoles.
It is interesting to try to understand 
the effect on the fields as one moves towards the boundary of  $Y$. 
The coefficient of $1/r$, $(r=|{\bf x}|),$ in
the asymptotic
expansion of the Higgs field, $\Phi({\bf x})$, of an SU(3) monopole is
$-\frac{i}{2}\mbox{diag}(-1,-1,2)$ whereas the 
$1/r$ coefficient of an embedded SU(2) monopole is 
$-\frac{i}{2}\mbox{diag}(-2,0,2)$. If one  is close
to the boundary of $Y$ then the fields should look like embedded SU(2)
fields (since the boundary corresponds to embedded SU(2) monopoles),
but with a cloud where the $1/r$ coefficient of the Higgs field changes
from -$\frac{i}{2}\mbox{diag}(-2,0,2)$ to
$-\frac{i}{2}\mbox{diag}(-1,-1,2)$ \cite{4}. 
A useful parametrization near the boundary of $Y$ which measures
approximately the radius of the cloud is given by
$z=2\sqrt{\frac{D}{\rho}}$ with $\rho=2K(k)-3D$: $z$ is related to the
inverse of the coordinate distance (not metric) from the boundary
of $Y$.
Near the boundary the expressions for the metric (A.14) may be
simplified to
\bea
ds^2 = dz^2+\frac{z^2}{4}(\hat{\sigma}_1^2
+\hat{\sigma}_2^2+\hat{\sigma}_3^2) 
+ \frac{b^2}{D^2}dD^2+a^2\sigma_1^2+b^2\sigma_2^2+c^2\sigma_3^2
\;\;+\;\;O(z^{-4})\;\;\;.
\eea
$\sigma_i,\,\hat{\sigma}_i$ are left-invariant 1-forms (defined in the
appendix (A.11)), corresponding to
rotational and SU(2) degrees of freedom;
$a^2,\,b^2,\,c^2$ are evaluated at the boundary ,$\;D=\frac{2}{3}K(k)$, giving
\be
a^2=\frac{2K(K-E)(E-k'^2K)}{3E}\,,\qquad
b^2=\frac{2EK(K-E)}{3(E-k'^2K)}\,,\qquad
c^2=\frac{2EK(E-k'^2K)}{3(K-E)}\,.
\ee
Here $k'^2=1-k^2$ and
$E(k)=\int_0^{\frac{\pi}{2}}(1-k^2\sin^2\theta)^{1/2}\,d\theta$.

Now, noting that (3.2) describes the flat space metric for $\R^4$ and 
the Atiyah-Hitchin (${\cal AH}$) metric for
SU(2) monopoles (with a scale factor of $1/3$) \cite{AH}, we 
see that in the limit the metric decouples into the direct product $\R^4\times
{\cal AH}$. We can interpret this by saying the monopole configuration
looks like an embedded charge two SU(2) monopole surrounded by a
cloud, parametrized by $\R^4$. In this limit the geodesics on
$M^8$ are easy to analyse. They are given by a straight line in the
$\R^4$ factor and the usual geodesics on the Atiyah-Hitchin manifold.
From numerical evidence in \cite{5}, it was shown that all geodesics
(except for the axially symmetric monopole collision) 
approach the asymptotic
regions of $Y$. In fact, we were able to check (using MATLAB) that
generic geodesics approach the boundary of the asymptotic region, 
i.e. $D\rightarrow \infty$ and $D\rightarrow
\frac{2}{3}K(k)$. It is interesting to ask whether similar behaviour
occurs for generic geodesics on $M^8$ not restricted to the
submanifold $Y$. From
(3.2) it can be seen that if a configuration is close to the boundary
and heading towards the boundary ($\dot{z}>0$) then the monopoles
will continue to approach $z=\infty$.
Also (3.2) suggests that the 
cloud will be spherically symmetric in the
limit $z\rightarrow \infty$ because the coefficients of $\hat{\sigma}_i^2$
are all equal.
  
With the assumption that near the boundary the long-range fields are 
spherically symmetric we may
now solve for these fields and independently verify that the $\R^4$
part of the metric in (3.2) is correct. This is simplest for the 
axially symmetric
trigonometric monopoles, ($k=0$).
For such monopoles in \cite{4}, the Higgs field was given along
the axis of axial symmetry. 
The long-range part of the Higgs field on the axis of symmetry may be 
found by
dropping all terms exponentially
small in $r$. Spherical symmetry determines
the long-range
Higgs field in all directions. On the axis of symmetry,

$\Phi=i$diag$(\Phi_{11},\,-\Phi_{11}-\Phi_{33},\,\Phi_{33})\,+O(e^{-6r})$ with
\be
\Phi_{11}
=-1+\frac{1}{D^2+4r^2}\left\{ \frac{(4r^2-D^2)\sin3D-4Dr\cos3D}
{2r\sin3D-D\cos3D}
\right\}\,,\;\;\Phi_{33}=2-\frac{4r}{4r^2+D^2}\;. 
\ee
To solve the spherically symmetric BPS equations
the Higgs field needs to be further truncated to  
\be
\Phi_{11}
=-1+\frac{1}{r}\left\{ \frac{r\sin3D-D\cos3D}{2r\sin3D-D\cos3D}
\right\}\;,\qquad\Phi_{33}=2-\frac{1}{r}\;. 
\ee
Because $D<\pi/3$ and $r$ is large this is a valid
approximation. The equations may then be solved 
for all values of
$D$. However we are only interested in configurations near the
boundary $(D\rightarrow \pi/3)$ where the $\R^4$ part of the metric decouples. 
We expand $\sin3D,\,\cos3D$ to order $\rho$, (here $\rho$ is $\pi-3D$
because $k=0$)                
and express
the fields in terms of $z=2\sqrt{\frac{D}{\rho}}$.
Then using \cite{6} the spherically symmetric 
BPS equations may be solved to give the long-range part of the gauge field 
$A_i^a$ (in a gauge with a Dirac string). The singularity at the origin
is not relevant here as we are only interested in the long-range
fields. We find
\bea
\Phi(r)&=&  i\left ( \begin{array}{ccc} -1 & 0 & 0 \\ 0 & -1 & 0 \\
0 & 0 & 2 \end{array} \right )  +
\frac{i}{r}\left ( \begin{array}{ccc} \frac{z^2+4r}{z^2+8r} &
0 & 0 \\ 0 & \frac{4r}{z^2+8r} & 0 \\ 0 & 0 & -1 \end{array} \right )  \\
{\bf A}(r,\theta)&=&\frac{i}{2r}\,\frac{8r}{z^2+8r}\left(\begin{array}{ccc} 
0 & -i & 0\\ i & 0 & 0\\ 0 & 0 & 0 \end{array}\right)\hat{\theta}
\\ \nonumber
&-&\left\{\frac{i}{2r}\,
\frac{8r}{z^2+8r}\left( \begin{array}{ccc} 0 & 1 & 0 \\ 1 & 0 & 0 \\ 0
& 0 & 0 \end{array} \right)  
-i\frac{\cos\theta}{r\sin\theta} 
 \left( \begin{array}{ccc} 1 & 0
& 0  \\ 0 & 0 & 0 \\ 0 & 0 & -1
\end{array} \right)\right\} \hat{\phi}
\eea
where $r$, $\theta$, $\phi$ are spherical polar coordinates.
If $z$ is large enough so that $z^2\gg r$   
then the $1/r$ coefficient in $\Phi$ is approximately $-\frac{i}{2}\mbox{diag}
(-2,0,2)$. 
However as $r\rightarrow\infty$ the $1/r$ coefficient in
$\Phi$ is $-\frac{i}{2}\mbox{diag}(-1,-1,2)$ as required.  
Near the boundary of $Y$ ($z\rightarrow \infty$), the change in 
fall-off in the 
$1/r$ term is extremely slow.
As $r\rightarrow \infty$ with $z^2\ll r,\;\Phi_{11}$ may be expanded
to give 
\be
\Phi_{11}=-i+\frac{i}{2r}+\frac{iz^2}{16r^2}\;,
\ee
so $z$ may be seen as changing the dipole term in $\Phi$. The
foregoing analysis leads us to interpret $z^2$ as the cloud radius.

From ${\bf A}$, the magnetic field ${\bf B}$ may be found, giving
\bea
{\bf B}&=&\frac{i}{2r^2}\left\{\left (
\begin{array}{ccc} 
-1 & 0 & 0 \\ 0 & -1 & 0 \\0 & 0 & 2 \end{array} \right )-
\frac{z^2(z^2+16r)}{(z^2+8r)^2}\left( \begin{array}{ccc} 1 & 0
& 0  \\ 0 & -1 & 0 \\ 0 & 0 & 0
\end{array} \right)
\right\}\hat{r} \\
&+&
\frac{4iz^2}{r(z^2+8r)^2}\left\{\left(\begin{array}{ccc} 
0 & 1 & 0\\ 1 & 0 & 0\\ 0 & 0 & 0 \end{array}\right)\hat{\theta}+
\left(\begin{array}{ccc} 
0 & -i & 0\\ i & 0 & 0\\ 0 & 0 & 0
\end{array}\right)\hat{\phi}\right\}\;\;.\nonumber
\eea
The magnetic field is seen to be non-Abelian in the cloud region
$r\approx z^2$. As $r\rightarrow \infty$ the non-Abelian components
decay like $1/r^3$ leaving an Abelian magnetic field at infinity,
\be
{\bf B}=\frac{i}{2}\left 
( \begin{array}{ccc} -1 & 0 & 0 \\ 0 & -1 & 0 \\
0 & 0 & 2 \end{array} \right )\,\frac{\hat{r}}{r^2}\;.
\ee 
The cloud may be
viewed as some form of shield for the non-Abelian magnetic field. The
fields described here exhibit very similar behaviour to the fields
given in \cite{LWY1} for the SO(5) monopole where again there is a
cloud parameter describing the extent to which the long-range
non-Abelian fields penetrate beyond the monopoles' core.

Now  $0\leq z<\infty$ and $k=0$ is part of a geodesic on $Y$ \cite{3}. 
The kinetic energy, $T$, from the asymptotic
fields obtained by varying $z$ may be calculated and compared to
that from the metric (the
constant of proportionality between the metric and the Lagrangian for
the two monopole system is one sixth the reduced mass of a pair of
monopoles, which is $\pi$). Normally one needs to add a
little gauge transform to ensure that the variation of the fields is
orthogonal to gauge orbits, but in this case, $D_i\delta A_i+[\Phi,
\delta \Phi]=0$ is satisfied anyway with
$\delta A_i=\dot{z}\,\partial A_i/\partial z\,\,$, 
$\delta \Phi=\dot{z}\,\partial \Phi/\partial z\;$. Thus, the kinetic 
energy $T$ is given by
\be
T=\frac{1}{2}\int d^3x\left\{<\frac{\partial \Phi}{\partial z},
\frac{\partial \Phi}{\partial z}> +
<\frac{\partial A_i}{\partial z},\frac{\partial A_i}{\partial z}>
 \right\}\;\dot{z}^2\;,
\ee
where $<X,Y>=-2\mbox{Tr}XY\;$. 
One finds
\be
T=\pi\dot{z}^2\;+\;O(z^{-4})\,\dot{z}^2\;.
\ee
In the
limit $z\rightarrow \infty$, this formula agrees with that found 
from the metric (3.2), however not to $O(z^{-4})$.
This means that as $z\rightarrow \infty$ all the kinetic energy is
outside the core (and indirectly verifies the assumption of asymptotic
spherical symmetry). 
The geodesic equation may be solved near the boundary
to give 
\be
z(t)=\beta t\;,
\ee
with $\beta$ constant and  total kinetic energy,
$T=\pi\beta^2$.

For the line of hyperbolic monopoles $(k=1)$, the Higgs field
was given on the axis of axial symmetry in \cite{4}. It was seen 
that there is no cloud in this case. The line $k=1$ (or $z=0$) is 
infinitely far
from the boundary and the cloud should not be expected to exist here.

The long-range fields for the 1-parameter family of trigonometric
monopoles is given in (3.6), (3.7). For all points in $Y$ near the
boundary the long-range fields must behave in a similar fashion, with
a cloud where the $1/r$ part of the Higgs field changes from
$-\frac{i}{2}$diag$(-2,0,2)$ to $-\frac{i}{2}$diag$(-1,-1,2)$. This cloud
will necessarily be larger than the monopoles' separation which is
$D$. We may view
the long-range fields as functions of $D$, $z$. 
In analogy with
the trigonometric case we expect that as $z\rightarrow\infty$ the 
kinetic energy of the long-range fields obtained by varying the cloud
parameter $z$ should equal the kinetic
energy derived from the corresponding term in the metric, i.e. $\pi
\dot{z}^2$.
This is indeed the case if the long-range fields are gauge equivalent
to (3.6), (3.7) for all $k$. 

It is worth noting that this cloud cannot be interpreted as a kink, i.e.
some region where the fields rapidly change their $1/r$ fall-off from 
$-\frac{i}{2}\mbox{diag}(-2,0,2)$ to $-\frac{i}{2}\mbox{diag}(-1,-1,2)$. 
As seen from (3.6) the
change in $1/r$ fall-off in $\Phi$ is very slow as
$z\rightarrow \infty$. Also the static energy density ${\cal{E}}$, $
({\cal{E}}=<B^{i},\,B^{i}>)$, in this region may be easily 
found from (3.9) to be
\be
{\cal E}=\frac{4}{r^4}-\frac{z^2+2r}{2r(r+z^2/8)^4}
\ee
We see that for large $z$, $\cal{E}$ is of order $1/r^4$ varying from
$4/r^4$ to $3/r^4$ 
as $r$ changes from $r\ll z^2$ to $r\gg z^2$. The energy density outside the
monopole cores is small irrespective of $z$. For large
$z$ the long-range fields and energy density
change only slightly with $z$.

At an arbitrary point in $M^8$ represented by $(\Phi,\,{\bf A})$ a
tangent vector may be written as $(\delta\Phi,\,\delta{\bf A})$ with 
$D_i\delta A_i+[\Phi,\delta \Phi]=0$, and $(\delta\Phi,\,\delta{\bf
A})$ also satisfy the linearized Bogomol'nyi equations. From one such
tangent vector other tangent vectors may be generated by 
\bea
\delta'\Phi&=&-\hat{\bf n}\cdot\delta{\bf A}\\
\delta'{\bf A}&=&\hat{\bf n}\delta \Phi+\hat{\bf n}\times\delta{\bf A}\nonumber
\eea
where $\hat{\bf n}$ is any constant unit vector. It is easy to see 
that these also
satisfy the above conditions for a tangent vector. By choosing an
orthonormal triplet $\hat{\bf n}_1,\,\hat{\bf n}_2,\,\hat{\bf n}_3$
we obtain four zero modes which are mutually orthogonal. Denoting by $J_i$
the action that takes $(\delta\Phi,\,\delta{\bf A})$ to 
$(-\hat{\bf n}_i\cdot\delta{\bf A},\,\hat{\bf n}_i\delta \Phi+\hat{\bf n}_i
\times\delta{\bf A})$
it is obvious that $J_iJ_j=-\delta_{ij}+\epsilon_{ijk}J_k$ thus
giving a realisation of the hyperk\"{a}hler structure on $M^8$.
For large $z$,$\;M^8$ splits as a product of hyperk\"{a}hler manifolds,
$\cal {AH}$ and $\R^4$. Thus one would expect that from the zero mode 
$\delta A_i=\delta z\,\partial A_i/\partial z$, 
$\,\delta \Phi=\delta z\,\partial \Phi/\partial z$
one may generate three other zero modes which correspond to the
unbroken SU(2) action. In fact, it is not difficult to check that
these zero modes may be written as
\bea
\delta' A_i&=&D_i\,\Delta({\bf r})\\
\delta' \Phi&=&[\Delta({\bf r}),\,\Phi]\nonumber
\eea
with $\Delta({\bf r})$ in the unbroken SU(2) whose explicit form is  
\bea
\Delta({\bf r})=\frac{8ir\delta z}{z(z^2+8r)}\left\{\hat{{\bf n}}
\cdot\hat{\theta}
\left(\begin{array}{ccc} 
0 & 1 & 0\\ 1 & 0 & 0\\ 0 & 0 & 0 \end{array}\right)+\hat{{\bf n}}
\cdot\hat{\phi}
\left(\begin{array}{ccc} 
0 & -i & 0\\ i & 0 & 0\\ 0 & 0 & 0 \end{array}\right)+\hat{{\bf n}}
\cdot\hat{r}\left(\begin{array}{ccc} 
1 & 0 & 0\\ 0 & -1 & 0\\ 0 & 0 & 0 \end{array}\right)\right\}\,.
\nonumber
\eea
The gauge rotation angle is given by
$\lim_{r\rightarrow \infty}\,\parallel\Delta({\bf r})\parallel$ 
which is $2\delta z/z$, 
independent of
$\hat{{\bf n}}$, where $\parallel X\parallel=\sqrt{<X,X>}\;$. 
This implies that the metric coefficients for the
SU(2) action are isotropic for large $z$ and their norm is $z^2/4$
times that of the $dz^2$ coefficient, in agreement with (3.2).  
\\[20pt]
\noindent{\bf Acknowledgments}
\news
\ \indent 
Thanks to Conor Houghton, Bernd Schroers and especially
Nick Manton for many helpful discussions. I also acknowledge the
financial support of the PPARC.
\\[20pt]

\appendix
\section{Appendix : The metric}
\setcounter {section}{1}
\setcounter {equation}{0}
\news
\ \indent

The natural metric on the moduli space is defined by computing the
$L^2$ norms of the zero-modes (solutions of the linearized Bogomol'nyi 
equations). The zero-modes must be orthogonal to little gauge
transforms, which means the zero-modes $\delta\Phi$,
$\delta{\bf A}_i$ must satisfy
\be
D_i\delta{\bf A}_i+[\Phi,\delta\Phi]=0
\ee
However in most cases it is extremely difficult to compute the metric
in this fashion and other techniques must be used
\cite{AH, Nahm}. Nahm's equations are the following system of
nonlinear ordinary differential equations
\bea
\dot{T}_1+[T_0,\,T_1]=[T_2,\,T_3]\\
\dot{T}_2+[T_0,\,T_2]=[T_3,\,T_1]\nonumber\\
\dot{T}_3+[T_0,\,T_3]=[T_1,\,T_2]\nonumber
\eea
where the $T_i$ are matrix-valued functions on an interval of the real
line parametrized by $s$. They are obtained from the self-dual 
Yang-Mills equations in
four dimensions by imposing translational invariance in three
dimensions just as the Bogomol'nyi equations are obtained by imposing
translational invariance in one dimension. It is known that solutions
to Nahm's equations with certain boundary conditions are in 1-1
correspondence with monopoles via the Atiyah-Drinfeld-Hitchin-Manin- 
(ADHM) Nahm transform. The
boundary conditions of Nahm's equations determine what gauge group the monopole
is in. This transform is known to be an isometry  for
SU(n+1) broken to U(n) \cite{NAK1, NAK2} (this includes 
SU(2) monopoles) and is believed to be true generally. 

Nahm data corresponding to Dancer monopoles are given by 
the space ${\cal C}$ of quadruples 
($T_0\,,T_1\,,T_2\,,T_3$) where

(i) $T_u$ ($u=0,\,1,\,2,\,3)$ are functions defined on the interval
$[0,\,3]$ taking values in the Lie algebra of U(2).

(ii) $T_0$ is analytic on $[0,\,3]$. $T_1$, $T_2$, $T_3$ are analytic on
$(0,\,3]$ with simple poles at $s=0$ of residue $\frac{i}{2}\tau_1$, 
$\frac{i}{2}\tau_2$, $\frac{i}{2}\tau_3$, where $\tau_i$ are the
Pauli matrices.

(iii) The $T_i$ satisfy Nahm's equations.

Nahm data for embedded SU(2) monopoles are as above except now the 
$T_i$ also have poles at $s=3$.
Let $G^0_0$ be the group of analytic U(2)-valued functions on $[0,3]$
which are the identity at $s=0,\,3$. This group acts on ${\cal C}$ in 
the following manner
\bea
T_0&\rightarrow& gT_0g^{-1}-\frac{dg}{ds}g^{-1}\\
T_i&\rightarrow& gT_ig^{-1}\;,\;\;\;\;\;\;\;\;\;(i=1,\,2,\,3)\;.\nonumber
\eea

The moduli space of Nahm data, denoted $M^{12}$, is the quotient of
${\cal C}$ by
$G^0_0$. $M^{12}$ has the following isometric group actions.

(a) There is an action of $\R^3$ on $M^{12}$, given by 
\be
T_i\rightarrow T_i-i\lambda_i\mbox{Id}\qquad(i=1,2,3)\,,
\ee
where $(\lambda_1,\lambda_2,\lambda_3)\in\R^3\,.$

(b) Let $G_0$ be the group of analytic U(2)-valued functions on $[0,3]$
which are the identity at $s=0$. This  acts on ${\cal C}$ as in
(A.3) with $g(3)=P$ say. This action descends to an action of 
$G_0/G_0^0\cong$U(2) on $M^{12}.$ 

(c) There is a Spin(3) action on $M^{12}$ induced from the
following action on ${\cal C}$.
Let $R\in$ SU(2) descend to $(R_{ij})\in$ SO(3). 
$g$ is an  analytic U(2)-valued function on $[0,3]$
which satisfies
$g(0)=R,\;g(3)=\,$Id. Then the Spin(3) action is defined by 
\bea
T_0&\rightarrow& gT_0g^{-1}-\frac{dg}{ds}g^{-1}\\
T_i&\rightarrow& g(\,\sum R_{ij}T_j\,)g^{-1}\;,\;\;\;\;\;\;\;\;\;
(i=1,\,2,\,3)\,.\nonumber
\eea
$M^8$ is defined as the quotient of $M^{12}$ by $\R^3\times$U(1)
where the U(1) is the center of U(2) and $N^5$ is the quotient
of $M^8$ by the SU(2) action. Using the SU(2) action to set $T_0$
to zero, Nahm's equations become 
$\dot{T}_1=[T_2,\,T_3]$ + cyclic. 
\newpage
We may now use the Spin(3) action to set $T_i=-\frac{i}{2}f_i\tau_i$ (no
sum on $i$) and Nahm's equations reduce to the Euler top equations
\be
\dot{f}_1=f_2f_3\;\;+\mbox{cyclic}
\ee
 whose solution is 
\be
f_1(s)=-\frac{Dcn_k(Ds)}{sn_k(Ds)}\,,\qquad  f_2(s)=-\frac{Ddn_k(Ds)}
{sn_k(Ds)}\,,\qquad   f_3(s)=-\frac{D}{sn_k(Ds)}\;.
\ee
$D$, $k$ are coordinates on the quotient space $N^5/$SO(3).
In order to express the metric it is useful to define $X,\,g_1,\,g_2$ by
\be
X(k,D)=f_1(3)f_2(3)f_3(3)\,,\qquad
g_1(k,D)=\int_0^3\frac{1}{f_2^2}\,ds\,,\qquad 
g_2(k,D)=\int_0^3\frac{1}{f_3^2}\,ds\;.
\ee
By the isometry proved in \cite{NAK2} 
the metric on the moduli space $M^{12}$ 
induced from the natural $L^2$ metric on ${\cal C}$
gives the metric on the moduli space of charge $(2,\,1)$ SU(3) monopoles.
In \cite{3}, an explicit expression was given for the metric on $N^5$
and an implicit expression for the metric on $M^8$. Here we want an
explicit expression for the metric in terms of $dD,\,dk$ and
left-invariant 1-forms corresponding to the SO(3)$\times$SO(3)
action. In \cite{3} the metric on
$N^5$ was expressed in terms of coordinates
$\alpha_i$ which
are related to the inner products of the Nahm matrices. Explicitly,
\bea
\alpha_1&=&<T_1,T_1>-<T_2,T_2>\;,\\
\alpha_2&=&<T_1,T_1>-<T_3,T_3>\;,\nonumber\\
\alpha_3&=&<T_1,T_2>\;,\nonumber\\
\alpha_4&=&<T_1,T_3>\;,\nonumber\\
\alpha_5&=&<T_2,T_3>\,,\nonumber
\eea
and again $<X,Y>=-2\mbox{Tr}XY$. 
$\alpha_i$ may be expressed in terms 
of $D$, $k$ and an SO(3)
matrix $(R_{ij})$. The SO(3) symmetry of the metric means that it
may be written in the form
\be
ds^2=A(k,D)dk^2 + 2B(k,D)dkdD + C(k,D)dD^2
+ a_{ij}(k,D)\sigma_i\sigma_j
\ee 
with $\sigma_i$ left-invariant 1-forms on SO(3) defined by
$\frac{i}{2}\tau_i\sigma_i=R^{\dagger}\,dR$ where $R\in$SU(2)
descends to $(R_{ij})$. If $\theta$,
$\phi$, $\psi$ are the usual Euler angles for SO(3), with
$0\,\leq\,\theta\,\leq\,\pi$, $0\,\leq\,\phi\,\leq\,2\pi$, 
$0\,\leq\,\psi\,\leq\,2\pi$,
then,
\bea
\sigma_1&=&-\sin\psi\,d\theta + \cos\psi\,\sin\theta\,d\phi\\
\sigma_2&=&\cos\psi\,d\theta + \sin\psi\,\sin\theta\,d\phi\nonumber\\
\sigma_3&=&d\psi+\cos\theta\,d\phi\;.\nonumber
\eea
The coefficients
$A, B, C$ are known from \cite{5}, however it is easier to
express the metric in terms of
$\tilde\alpha_1=-(1-k^2)D^2,\;\tilde\alpha_2=-D^2$, 
which are
$\alpha_1,\,\alpha_2$ restricted to $N^5/$SO(3).
$a_{ij}$ can be found by letting 
$R= 1+\delta$ with all entries in $\delta$ close to zero and
expressing the metric in terms of the invariant
1-forms at this point. Hence (A.10) may be written as 
\be
ds^2=\frac{1}{4}\left\{X(g_1d\tilde\alpha_1+g_2d\tilde\alpha_2)^2
+ g_1d\tilde\alpha_1^2 + g_2d\tilde\alpha_2^2\right\}+
a_1\sigma_1^2+a_2\sigma_2^2+a_3\sigma_3^2
\ee
with 
\bea
a_1&=&k^4D^4\frac{g_1g_2}{g_1+g_2} \\
a_2&=&D^4\left\{g_2+\frac{Xg_2^2}{Xg_1+1}\right\}\nonumber \\
a_3&=&D^4(1-k^2)^2\left\{g_1+\frac{Xg_1^2}{Xg_2+1}\right\}\;.\nonumber 
\eea
$a_{ij}$ is diagonal due
to a $\Z_2 \times \Z_2$ symmetry on $N^5$. The coefficients of the
metric depend only on $D$, $k$.
The metric on $M^8$ will contain the left-invariant 1-forms
$\sigma_i,\check{\sigma}_i$, corresponding to the SO(3)$\times$SO(3)
action where $\check{\sigma}_i$ is defined by
$\frac{i}{2}\tau_i\check{\sigma}_i=P^{\dagger}\,dP$ and again $P\in$
SU(2) descends to $(P_{ij})\in$ SO(3). 
$\hat{\sigma}_i$ which appeared in Section $3$ will be defined
below in terms left $and$ right-invariant 1-forms corresponding to $P,\,R$.
The metric coefficients on $M^8$ in \cite{3} are determined by
orthogonality conditions depending on $D$, $k$ and $(R_{ij})$, $(P_{ij})$ in
SO(3).
These may be solved for $(R_{ij})$, $(P_{ij})$ in the neighbourhood 
of the identity to
get the metric in this region in terms of $d\tilde\alpha_1$, 
$d\tilde\alpha_2$, $\sigma_i,\check{\sigma}_i$
and again this holds everywhere in $M^8$ due to the SO(3)$\times$SO(3) 
symmetry of the metric. We have
\bea
ds^2&=&\frac{1}{4}\left\{X(g_1d\tilde\alpha_1+g_2d\tilde\alpha_2)^2
+ g_1d\tilde\alpha_1^2 + g_2d\tilde\alpha_2^2\right\}+
a_1\sigma_1^2+a_2\sigma_2^2+a_3\sigma_3^2\\
&+&\left\{b_1\sigma_1-c_1(\frac{f_3(3)}{f_2(3)}\sigma_1-
\check{\sigma}_1)
\right\}^2+\left\{b_2\sigma_2-c_2(\frac{f_1(3)}{f_3(3)}\sigma_2-\check
{\sigma}_2)\right\}^2\nonumber\\
&+&\left\{b_3\sigma_3-c_3(\frac{f_1(3)}{f_2(3)}\sigma_3-\check{\sigma}_3)
\right\}^2
\nonumber
\eea
$a_i$ are as before, and 
\bea
b_1&=&k^2D^2\sqrt{\frac{g_1^2}{(g_1+g_2)
(Xg_1+Xg_2+1)}}\;,\quad
c_1=f_2(3)f_3(3)\sqrt{\frac{g_1+g_2}{Xg_1+Xg_2+1}},\\
b_2&=&\frac{g_2D^2}{\sqrt{Xg_1^2+g_1}}\;, \quad\qquad\qquad\qquad\qquad\quad\;\,
c_2=\frac{\sqrt{Xg_1^2+g_1}}{f_2(3)g_1}\;,\nonumber\\
b_3&=&\frac{g_1D^2k'^2}{\sqrt{Xg_2^2+g_2}}\;, \quad\qquad\qquad\qquad\qquad\quad\;\,
c_3=\frac{\sqrt{Xg_2^2+g_2}}{f_3(3)g_2}\;.\nonumber
\eea
Again $b_i$, $c_i$ depend only on $D$, $k$.
$D$ satisfies $0\leq D<\frac{2}{3}K(k)$ . The manifold thus has a boundary 
$D=\frac{2}{3}K(k)$ corresponding to Nahm data which are singular at 
$s=0$ and $s=3$.
As one approaches the boundary on $M^8$ the metric (A.14) becomes
\bea
ds^2 &=& dz^2+\frac{z^2}{4}\left\{(\sigma_1-\check{\sigma}_1)^2
+(\sigma_2+\check{\sigma}_2)^2+(\sigma_3+\check{\sigma}_3)^2\right\}\\ 
&+&
\frac{b^2}{D^2}dD^2+a^2\sigma_1^2+b^2\sigma_2^2+c^2\sigma_3^2\;+\;O(z^{-4})\;,
\nonumber
\eea
with $a^2,\,b^2,\,c^2$ defined in (3.3). As stated previously the
second line in (A.16) is the Atiyah-Hitchin metric. We now do a
coordinate transform so that the first line in (A.16) is explicitly
seen to be the flat metric on $\R^4$.
We redefine 
the Spin(3) action, (A.5), so that 
$g(3)=e^{\frac{i\pi}{2}\sigma_1}R^{\dagger}e^{\frac{i\pi}{2}\sigma_1}$
(the purpose of the $e^{\frac{i\pi}{2}\sigma_1}$ terms is to change the
signs of $\check{\sigma}_2,\,\check{\sigma}_3)$.
$\hat{\sigma}_i$ which appears in (3.2) is defined by 
$\frac{i}{2}\tau_i\hat{\sigma}_i=X^{\dagger}dX$ where 
$X=Pe^{\frac{i\pi}{2}\sigma_1}R^{\dagger}e^{\frac{i\pi}{2}\sigma_1}$.
Expressed in terms of 1-forms corresponding to the matrices $R,\,P$
the metric will now contain both 
left and right-invariant 1-forms but for large
cloud parameter (A.16) now takes the simple form given in (3.2),
essentially because $\sigma_1^2+\sigma_2^2+\sigma_3^2$ is independent
of whether $\sigma_i$ is a left or right-invariant 1-form.  
The point is
that $g(3)$ can be chosen arbitrarily and we choose it as above so
that the form of the metric near the boundary, (A.16), takes an especially
simple form. 
\\[20 pt]


\begin{thebibliography}{99}
\bibitem{3} A.S. Dancer, Commun. Math. Phys. 158 (1993) 545.
\bibitem{4}  A.S. Dancer, Nonlinearity 5 (1992) 1355.
\bibitem{5}  A.S. Dancer and R.A. Leese, Proc. R. Soc. 440 (1993) 421.
\bibitem{13}  A.S. Dancer and R.A. Leese, Phys. Lett. B390 (1997) 252.
\bibitem{GL} J. Gauntlett and D. Lowe, Nucl. Phys. B472 (1996) 194.

K. Lee, E. J. Weinberg, P. Yi, Phys. Lett. B376 (1996) 97.
\bibitem{LWY1} K. Lee, E. J. Weinberg, P. Yi, Phys. Rev. D54 (1996) 6351.
\bibitem{Connell} S.A. Connell, The dynamics of the SU(3) charge
(1,1) magnetic monopole, University of South Australia preprint.
\bibitem{LWY3} K. Lee, E. J. Weinberg, P. Yi, Phys. Rev. D54 (1996) 1633.
\bibitem{GW} G.W. Gibbons and P. Rychenkova, hep-th/9608085, to appear
in Commun. Math. Phys.
\bibitem{GNO} P. Goddard, J. Nuyts and D. Olive, Nucl. Phys. B125
(1977) 1.
\bibitem{19} A. Abouelsaood, Phys. Lett. 125B (1983) 467.
\bibitem{20} A. Abouelsaood, Nucl. Phys. B226 (1983) 309.
\bibitem{21} P. Nelson and A. Manohar, Phys. Rev. Lett. 50 (1983) 943.

A. Balachandran, G. Marmo, M. Mukunda, J. Nilsson, E. Sudarshan and
F. Zaccaria, Phys. Rev. Lett. 50 (1983) 1553.
\bibitem{Bow} M.C. Bowman, Phys. Rev. D32 (1985) 1569.
\bibitem{Murray} M. K. Murray, Comm. Math. Phys 125 (1989) 661.
\bibitem{Wein1}  E. J. Weinberg, Nucl. Phys. B167 (1980) 500.
\bibitem{Nahm} W. Nahm, in Monopoles in Quantum Field
Theory, N. Craigie et al. (eds.) Singapore: World Scientific (1982). 
\bibitem{NAK1} H. Nakajima, in Sanda 1990, Proceeedings, Einstein 
metrics and Yang-Mills connections. 
\bibitem{NAK2} M. Takahasi, Phd. Thesis, University of Tokyo.
\bibitem{EW} F. Englert and P. Windey, Phys. Rev. D14 (1976) 2728.
\bibitem{Bais} F. A. Bais, Phys. Rev. D18 (1978) 1206.
\bibitem{Hurt} J. Hurtubise, Commum. Math. Phys 120 (1989) 613.
\bibitem{Chalmers} G. Chalmers, Multi Monopole Moduli Spaces for $SU(N)$ Gauge
Group, hep-th/9605182.
\bibitem{BW} F. A. Bais and D. Wilkinson, Phys. Rev. D19 (1979) 2410. 
\bibitem{AH} M. F. Atiyah and N. J. Hitchin, The geometry and dynamics
of magnetic monopoles, Princeton University Press 1988.
\bibitem{6}  N.S. Manton, Ann. Phys. 132 (1981) 108.
\end{thebibliography}
\end{document}